\documentclass[a4paper,fleqn,usenatbib]{mnras}

\usepackage{newtxtext,newtxmath}

\usepackage[T1]{fontenc}
\usepackage{ae,aecompl}

\usepackage{graphicx}	
\usepackage{amsmath}	
\usepackage{tablefootnote}

\title[Turbulent component of solar dynamo]{Signature of the turbulent component of solar dynamo on active region scales and its association with flaring activity }

\author[V.I.Abramenko]{
Valentina I. Abramenko\thanks{E-mail: vabramenko@gmail.com (VIA)}
\\
Crimean Astrophysical Observatory, Russian Academy of Science, Nauchny, Bakhchisaray,  298409, Crimea, Russia
}

\date{Accepted XXX. Received YYY; in original form ZZZ}

\pubyear{2015}

\begin{document}
\label{firstpage}
\pagerange{\pageref{firstpage}--\pageref{lastpage}}
\maketitle

\begin{abstract}

Obtaining observational evidence of the turbulent component of solar dynamo operating in the convective zone is a challenging problem because the dynamo action is hidden below the photosphere. Here we present results of a statistical study of flaring active regions (ARs) that produced strong solar flares of an X-ray class X1.0 and higher during a time period that covered solar cycles 23 and 24. We introduced a magneto-morphological classification (MMC) of ARs, which allowed us to estimate possible contribution of the turbulent component of the dynamo into the structure of an AR. We found that in 72\% of cases, flaring ARs do not comply with the empirical laws of the global dynamo (frequently they are not bipolar ARs, or, if they are, they violate either  Hale polarity law, or the Joy's law, or the leading sunspot prevalence rule), which may be attributed to the influence of the turbulent dynamo action inside the convective zone on spatial scales of typical ARs. Thus it appears that the flaring is governed by the turbulent component of solar dynamo.  A contribution into the flaring from these  ARs-``violators'' is enhanced during the second maximum and the descending phase of a solar cycle, when the toroidal field weakens and the influence of the turbulent component becomes more pronounced. These observational findings are in consensus with a concept of the essential role of non-linearities and turbulent intermittency in the magnetic fields generation inside the convective zone, which follows from simulations of dynamo.

\end{abstract}

\begin{keywords}
Sun:magnetic fields -- Sun:photosphere -- turbulence -- diffusion
\end{keywords}



\section{Introduction}

Improved astrophysical observational and computational capabilities presented new evidence that magnetic fields are the key force that supports the endless chain of cosmological non-stationary phenomena. In space, the magnetic field generates energy from stars and galaxies. This small amount of energy under the influence of the weak seed magnetic field and turbulent motions in the medium is spent on generation of
new magnetic flux, the so called dynamo process. Release of this magnetic energy fuels the relentless activity of solar-type stars - spots and eruptions (flares on the Sun). 

Studies of the solar dynamo and investigations in the field of solar flare forecast usually do not overlap. Theoretical research and numerical simulations of the dynamo \citep[e.g.,][]{Karak2017,Pipin2018,Cameron2018}, see also reviews by \citet{Charbonneau2010,Charbonneau2014,Charbonneau2020,Brun2015}, do not consider problems of solar flares. In turn, recent progress in flare forecasting \citep[e.g.,][]{Barnes2016,Leka2019,Cinto2020,Nishizuka2020} is often based on photospheric magnetic field properties of active regions with statistical and machine learning techniques and does not involve processes of magnetic flux generation. 
Here we will explore solar activity via studying dynamo-processes and we will show that the non-linear (turbulent) component of the dynamo, namely, fluctuations of dynamo on a broad range of spatial scales, is connected to variations of flaring activity and they (fluctuations) can be revealed from observations.   
It is widely accepted that for the majority of solar active regions (ARs) the magnetic field is generated by the mean-field dynamo (global dynamo), see, e.g., a review by \citet{Lidia2015}. According to \citet{Abramenko2018}, the magnetic structure of about 70\% of the investigated 1494 ARs is consistent with the essential empirical laws that follow from the mean-field dynamo theory: bipolar ARs obeying the Hale polarity law, the Joy's law and the rule of a prevalence of the leading spot in a bipolar structure. Nevertheless, a question how to explain appearance of those 30\% of ARs that violate the aforementioned laws is still open. It is highly unlikely that these ARs are simply due to large fluctuations in an Gaussian medium since a fraction of such fluctuations should not exceed 5\%. One possibility is to consider them as a result of strong fluctuations in an intermittent medium, in other words, in a non-linear dynamical dissipative system (NLDDS). In such a system, strong fluctuations are not rare and they appear in both space and time domains. In this case, there is a basis to speculate that strong flares are intrinsically related to strong spatial fluctuations, i.e, ARs that violate the mean-field theory rules mentioned above.
It is well established that ARs with a complex magnetic structure display enhanced flare activity 
\citep[e.g.,][]{Ireland2008,Falconer2008,McAteer2010}.
However, there also are alternative opinions on this subject \citep[e.g.,][]{Georgoulis2012}. The majority of publications are focused on revealing of critical conditions for flares to occur, and on finding a set of parameters, which is sensitive to the pre-flare energy build-up. Our approach is different. We consider strong flares as unavoidable strong fluctuations in the time domain. For an NLDDS, the presence of strong temporal fluctuations implies the existance of strong deviations in the spatial domain as well. In our case, strong spatial fluctuations are the ``violator'' ARs with deviations from the regular magnetic configuration.  If the solar dynamo performs as a NLDDS, then occurrence of strong flares and the appearance of AR-``violators'' (irregular ARs) have to be statistically related. We intend to check this hypothesis. To do this, we investigated all 79 ARs of 23rd and 24th solar cycles that produced at least one X-class flare during its passage across the solar disk. For each AR we determined their magneto-morphological class (MMC) by using the criteria outlined below, and compared these with the flaring index of the AR. A time distribution of the ARs through the cycles was also analyzed to reveal the signature of the dynamo wave performance.
\section{Magneto-morphological classification of active regions}
According to the mean-field dynamo theory, flux tubes of the toroidal field rise from the bottom of the convection zone toward the photosphere and then form bipolar magnetic regions (sunspot groups, or ARs) as that flux pushes further into the solar atmosphere. For the majority of cases, the sunspot groups obey certain empirical laws \citep{Lidia2015}. Thus, polarities of the leading (western) sunspot are opposite in the Northern and Southern hemispheres, and the polarity sign changes from one cycle to another cycle, a behaviour known as Hale's polarity law \citep{Hale1919,Hale1925}.
Bipolar ARs tend to emerge with a systematic tilt of their axis relative to the solar equator, so that the leading sunspot is located closer to the equator. The tilt tends to increase with the latitude and this pattern is known as Joy's law. The twist of an emerging flux tube is determined by Coriolis force and is thought to be a plausible reason for the tilt \citep{WangSheeley1991,DSilva1993}, however other reasons are possible \citep{Leighton1969,McClintock2013}. 
The third empirical law is the prevalence of the leading sunspot in the bipolar structure: in majority of cases, the area of the leading spot is larger than the area of the largest sunspot in the following part of an AR. This implies that the leading spot is more coherent and the magnetic fields are less inclined than those in the following part. Note that \citet{Babcock1961} considered this observational property of bipolar ARs as one of the keystones of Babcock-Leighton phenomenological concept of the solar dynamo, that later became the origin of the mean-field dynamo theory.

However, as it was mentioned in Introduction, about one third of ARs do not follow these laws \citep{Abramenko2018}. The tilt of a bipolar structure may not follow Joy's law (the most common violation, 20\% of all ARs), the leading spot may be smaller than the largest following spot (about 12\% of the total number), or the polarities of leading and following spots may be reversed (anti-Hale ARs, they constitute about 3-4\% of all ARs). All these deviations may be explained by peculiarities in the flow field of the convection zone that twist and stretch a toroidal flux rope while it rises toward the photosphere, in other words, by the mild influence of the turbulent dynamo. 

The influence of turbulent dynamo on the toroidal flux ropes can be stronger. For example, it can result in fragmentation of a flux rope with subsequent deformation of the fragments that may lead to formation of several co-aligned bipoles on the solar surface with prevailing E-W orientation, similar to the well known AR NOAA 11158. Moreover, distortions of the toroidal flux rope may become so significant that the resulting AR may appear as a complex configuration of mixed polarity sunspots distributed chaotically.
 
In our original magneto-morphological classification \citep[MMC,][]{Abramenko2018}, we divided all ARs into three classes: class A, which includes bipolar ARs that follow all aforementioned laws, is referred to as regular ARs; class U consists of unipolar sunspots without opposite polarity pores in the trailing part, and class B that includes the rest of ARs - irregular ARs. Some of irregular ARs have a bipole structure and violate at least one of the laws. However, a part of them do not display a classical bipole structure; instead they show multipolar or strong $\delta$-structures. The multipolar and strong $\delta$-structures  were very rare when we studied ARs of any flaring capability, but they became not rare at all when we explored here the strong-flaring ARs. Strictly speaking, the three aforementioned laws are not applicable for ARs without classical bipolar structure. This motivated us to introduce here further specifications inside classes.

\begin{figure*}
 	\includegraphics[width=\columnwidth]{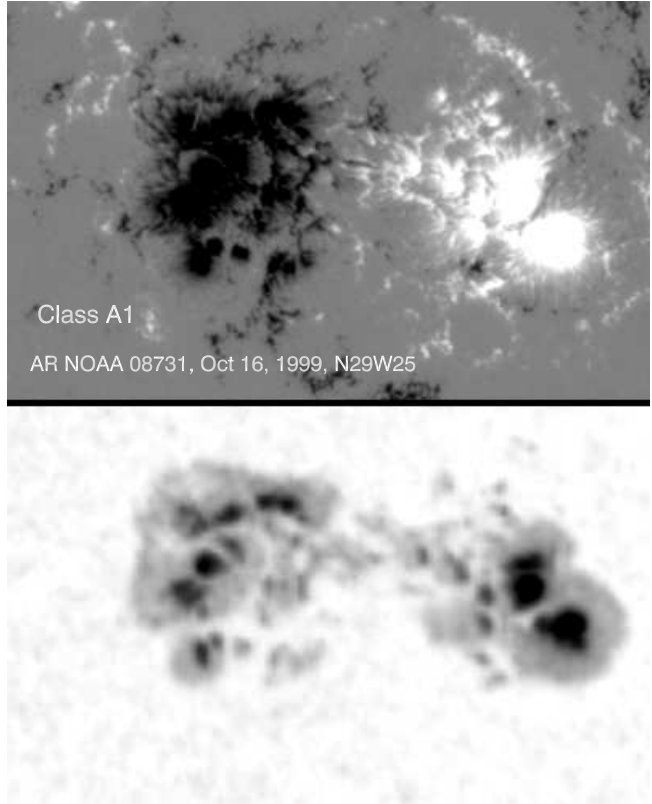}
 	\caption{\sf A typical example of an active region of A1-class: a bipolar AR of the 23rd cycle is located in the Northern hemisphere and has the positive polarity of the leading spot (compliance of the Hale polarity law); the AR obeys the Joy's law; the leading spot dominates any of the following spots. No small $\delta$-structure is observed. The magnetogram (top) and the continuum image (bottom) are acquired by SOHO/MDI instrument in the high-resolution mode \citep{Scherrer}.  North to the top, west to the right.  Direction of the equator coincides with the horizontal side of the frame. The magnetogram is scaled from -800~G (black) to 800~G (white). }
 	\label{fig-1}  
 \end{figure*}

\begin{figure*}
	\includegraphics[width=\columnwidth]{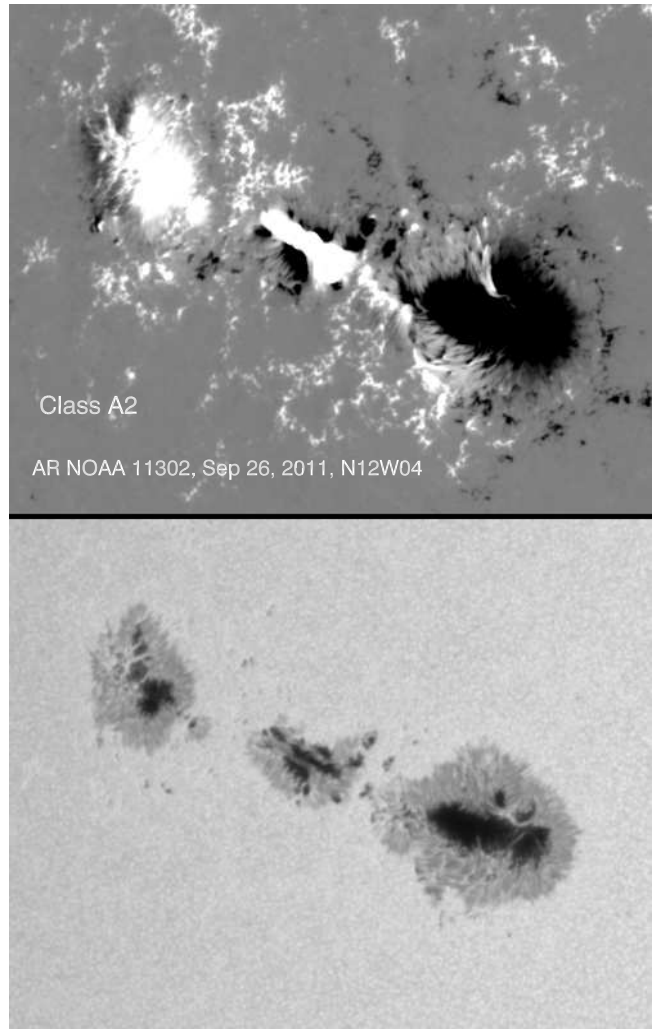}
	\caption{\sf  A typical example of an active region of A2-class: a bipolar AR of the 24rd cycle is located in the Northern hemisphere and has the negative polarity of the leading spot (compliance of the Hale polarity law); the AR obeys the Joy's law; the leading spot dominates the following spot. A small $\delta$-structure is observed in the middle. 
	The line-of-sight magnetogram (hmi.sharp-720s series) and the continuum image are acquired by the Helioseismic and Magnetic Imager (HMI) onboard  Solar Dynamic Observatory (SDO, \citet{Schou2012}). North to the top, west to the right.  Direction of the equator coincides with the horizontal side of the frame. The magnetogram is scaled from -500~G (black) to 500~G (white). 
  }
	\label{fig-2}  
\end{figure*}

\begin{figure*}
	\includegraphics[width=\columnwidth]{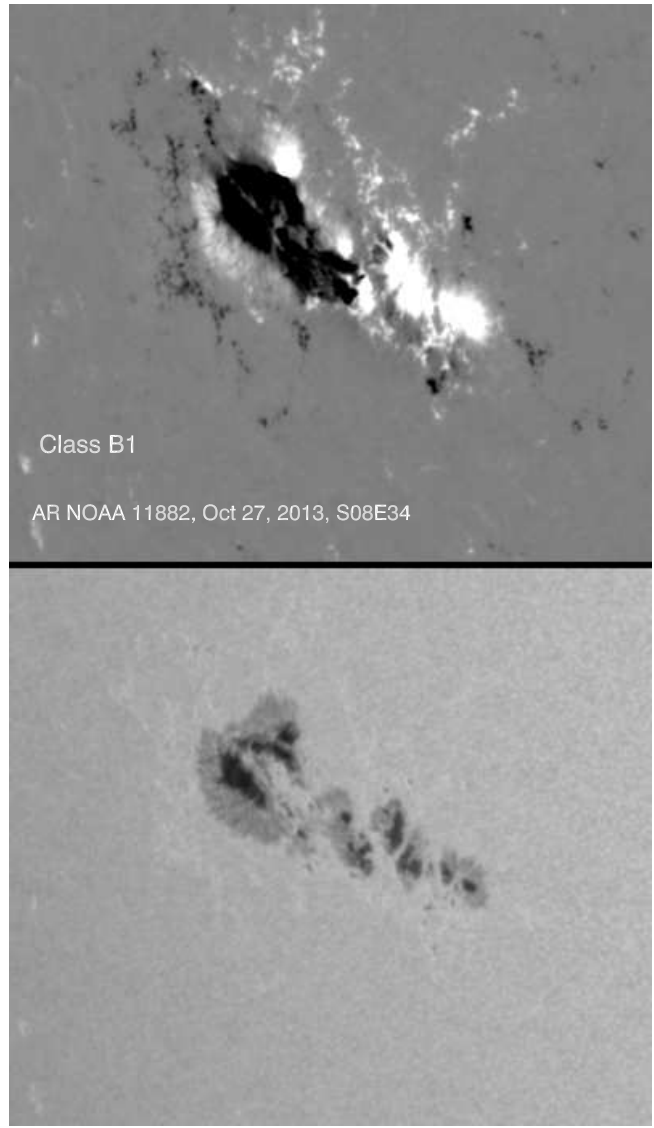}
	\caption{\sf  A typical example of an active region of B1-class: a bipolar AR of the 24rd cycle is located in the Southern hemisphere and has the positive polarity of the leading spot (compliance of the Hale polarity law); the AR does not obey the Joy's law: the leading part is located farther from the equator than the following part; the rule of the prevalence of the lading spot is not met: the leading spot is smaller than the following spot.  
	Notations are the same as in Figure \ref{fig-2}. 
  }
	\label{fig-3}  
\end{figure*}

\begin{figure*}
	\includegraphics[width=\columnwidth]{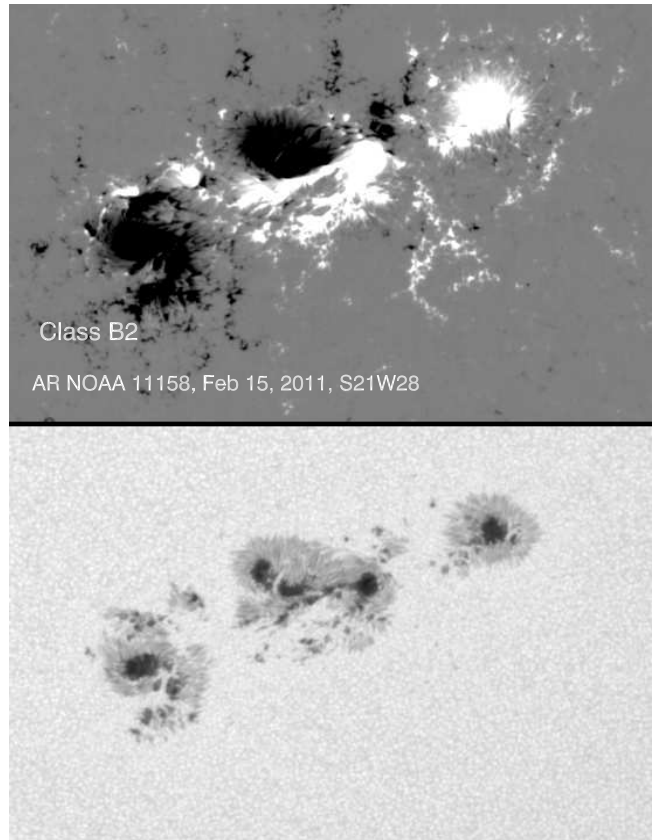}
	\caption{\sf An example of an active region of B2-class: a complex AR composed from two co-aligned bipoles emerged simultaneously with the general orientation in accordance with the Joy's law that allows to suppose a common toroidal flux tube (fragmented or bended, see Figure 6 in \citet{Toriumi2017}.)
	The Hale polarity law and the Joy's law are met for both bipoles, however the leading spot prevalence rule is not applicable. 	
	Notations are the same as in Figure \ref{fig-2}. 	
  }
	\label{fig-4}  
\end{figure*}

\begin{figure*}
	\includegraphics[width=\columnwidth]{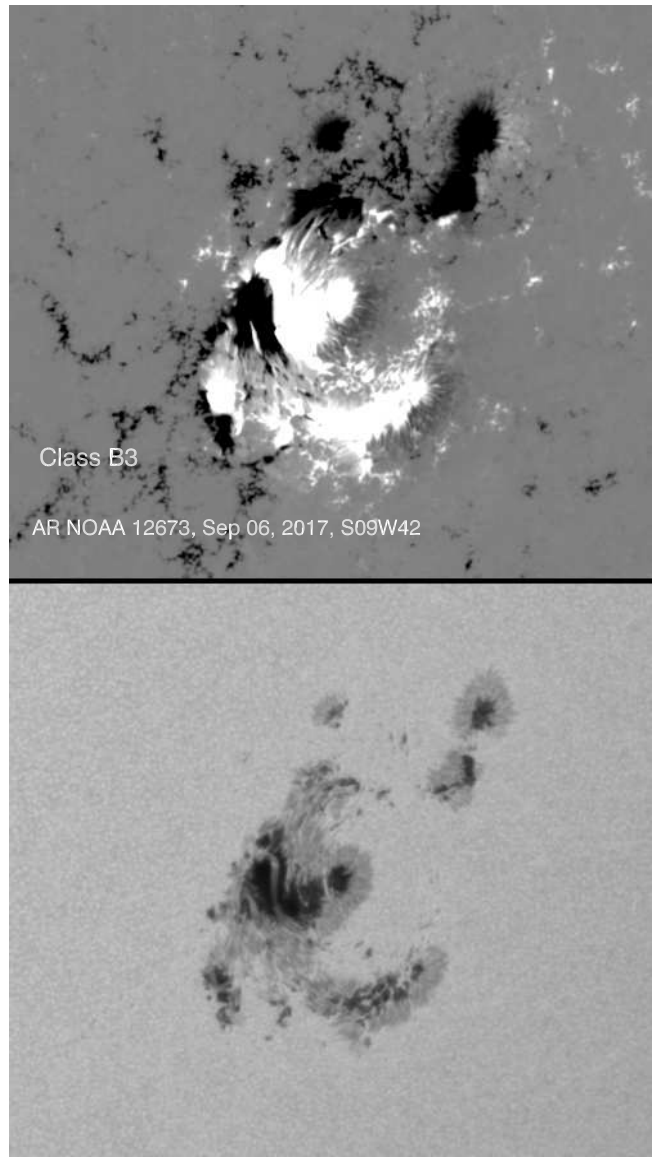}
	\caption{\sf An example of an active region of B3-class: a complex AR composed from several chaotically distributed spots of both polarities. The empirical global dynamo laws are not applicable. A complex knot of flux tubes in the convective zone, similar to that in Figure 4 by \citet{Ishii1998}, might be the source. 
	Notations are the same as in Figure \ref{fig-2}. 
}
	\label{fig-5}  
\end{figure*}

Considering that the aim of this study is to estimate the degree of influence of the turbulent dynamo on the AR formation process, we modified here our classification \citep{Abramenko2018}  by introducing sub-classes into the A and B classes:

\begin{description}	
\item[A1] - bipolar ARs for which Hale polarity law and Joy's law are fulfilled, the leading spot is dominant, and there were no small $\delta$-structures inside the AR during its passage across the solar disk (see Figure \ref{fig-1}). These type of ARs may be considered as a result of non-disturbed emergence of a single toroidal flux tube.
\item[A2] -  bipolar ARs for which Hale polarity law, the Joy's law are fulfilled, the leading spot dominates and there were small (relative to the size of the leading spot) $\delta$-structure(s) during the passage across the solar disk (see Figure \ref{fig-2}). These type of ARs may result from a small contribution of the turbulent dynamo, possibly operating at the near-surface depth.
\item[B1] - bipolar ARs which violate at least one of the aforementioned law (can be considered as a result of mild distortion of a single toroidal flux tube),  with small (if any) $\delta$-structure(s) during the passage across the solar disk.  A typical example is shown in Figure \ref{fig-3}. 
\item[B2] - multipolar ARs consisting of several quasi-coaligned bipoles having general axis orientation in accordance to Joy's law. Frequently these ARs contain a strong  $\delta$-structure and they can be represented by AR NOAA 11158 (Figure \ref{fig-4}). Class B2 ARs may be regarded as resulting of fragmentation and distortion of a single toroidal flux tube. For that reason we included into the B2 class strong single $\delta$-structures \citep[see Figure 15a in][]{Toriumi2019} as also consisting of one flux tube.
\item[B3] - multipolar ARs where opposite polarity sunspots are distributed in an irregular manner so that it is impossible to define the AR axis and assign leading and trailing sunspots (Figure \ref{fig-5}). These ARs represent the most complex magnetic structures and can be considered as a result of interaction (intertwining) of several flux tubes in the convective zone. Often such magnetic knots appear in the vicinity of a unipolar sunspot (e.g., AR NOAA 12673) resulting from fast flux emergence. It is likely that such a bundle of interwound flux tubes may have been channelled to the surface along the pre-existing structures of deeply rooted stable sunspots. The existence of such a channel in the vicinity of a vortex structure (recall that $B=rot(A)$) is a frequent occurrence in an intermittent medium \citep{Frisch1995}. 
\end{description}

 Table \ref{tab1} summarizes the essential criteria of the magneto-morphological classification. In parentheses, our comments in the framework of turbulent dynamo influence are presented.

\begin{table*}
	\centering
	\caption{Magneto-morphological classification of all ARs except unipolar sunspots. }
	\label{tab1}
	\begin{tabular}{p{2.5cm} p{2.5cm}|c|p{2.5cm} p{2.5cm} p{2.5cm}} 
		\hline
		\multicolumn{2}{ |c| }{A} & &\multicolumn{3}{ |c| }{B}\\
		\multicolumn{2}{ |l| }{Regular ARs:} & &\multicolumn{3}{ |l| }{All the rest: Irregular ARs}\\
		\multicolumn{2}{ |l| }{bipolar ARs obeying Hale polarity law, Joy's law, Leader prevalence rule } & &\multicolumn{3}{ |l| }{ }\\
		\cline{1-2} \cline{4-6}
		A1 & A2 && B1 & B2 & B3\\
		\hline
		Bipolar ARs obeying the Hale, Joy's laws, Leader prevalence rule, without any $\delta$-structures. (Emergence of a single toroidal flux tube following the global dynamo rules).&
		Bipolar ARs obeying the Hale, Joy's laws, Leader prevalence rule, with small $\delta$-structure(s). (Emergence of a single toroidal flux tube following the global dynamo and minor influence of turbulent dynamo). &&
		Bipolar ARs violating at least one of the laws. (Emergence of a single toroidal flux tube rotated and/or inclined owing to the turbulent dynamo action).&
		Multipolar ARs consisting of several co-aligned bipoles (as a result of fragmentation and distortion of a single flux tube), or tight strong  $\delta$-structure (as a result of strong twist of a single flux tube by the turbulent dynamo action). &
		Multipolar ARs with chaotically distributed spots of both polarities. (Emergence of several interwound flux tubes by turbulent dynamo).\\
		\hline
	\end{tabular}
\end{table*}

We note that some ARs evolved very fast during their passage across the solar disk, so that it was difficult to assign a permanent MMC class for the entire time interval. In such cases, we assigned the class as determined during a 2-3 day interval prior to the strongest flare. For ARs with the strongest flare occurred in the eastern limb, the class was acquired when the AR was on the longitude around -(50-70) degrees (e.g., NOAA ARs 12339, 10930).

As for bipolar ARs, in the present study, we utilized an experience obtained in \citet{Abramenko2018} for reliable estimations of the AR's tilt (Joy's law) and the leading spot dominance. Problems that arise in detection of ARs with reverse polarity (violation of the Hale polarity law) are discussed in details in \citet{Zhukova2020} and taken into account here. As for multipolar ARs, there could be some issues to which class (B2 or B3) a given AR should be assigned. For example, in NOAA AR 12297 the dominating feature is the strong $\delta$-structure, and the ARs could be classified as B2. However, a moderate bipole nearby the $\delta$-structure emerging during the day of the strongest flare is in favour to classify this AR as B3 - a multipolar caused by at least two flux tubes.

Our list includes 79 ARs that produced X-class flares during 
the time interval from January 1996 to December 2018 spanning solar cycles 23 and 24 (Tables \ref{tab2}, \ref{tab3}). Only for 9 ARs (out of 79) no $\delta$-structure were observed and documented during all days of observations (see the last column in tables). This allowed us to conclude that the presence of a $\delta$-structure within an AR appears to be a common condition for X-class flare to occur. 

In this context, it is interesting to compare our classification with the classifications of $\delta$-structures by \citet{Zirin1987} and by \citet{Toriumi2019}. Two opposite polarity umbras embedded in a common penumbra belong to type 1 according to \citet{Zirin1987} and comprises a``spot-spot'' type according to \citet{Toriumi2019}, while in our MMC classification this structure belongs to class B2. A multi-polar complex of tightly packed sunspots within an extended penumbra (``island $\delta$-spots'') belongs to type 1 $\delta$-structures per \citet{Zirin1987} and B3 class in the MMC classification. ARs of type 2 \citep{Zirin1987} and ``spot-satellite'' ARs \citep{Toriumi2019}, in cases when the satellite-sunspot is smaller than the leading spot, belong to the class of the hosting bipole. Type 3 ARs \citep{Zirin1987} and ``quadrupole'' groups \citep{Toriumi2019} overlap with class B2 in MMC classification. ``Inter-AR'' groups \citep{Toriumi2019} are very rare and belong to our B3-class (AR NOAA 08647, marked with a star in Table \ref{tab2}). 

We did not utilize the existing classifications such as Zurich classification (or McIntosh Sunspot Group Classification \citet{McIntosh1990}), Hale classification (Mount Wilson classification \citet{Hale1919}) for the following reasons. Both  these classifications uniformly treat all bipolar ARs, which is not acceptable when one aims to diagnose turbulent dynamo in the convection zone. The advantage of the proposed classification is that 1) those bipoles violating the empirical laws of the global dynamo are separated into one special class, and 2) the classification scheme is organized in such a way that the expected contribution from the turbulent dynamo increases through the classes from A1 toward B3.

\section{Distribution of strong-flaring ARs over the magneto-morphological classes}

Second and third columns in Tables \ref{tab2} and \ref{tab3} list the first and the last day of AR's presence on the solar disk, while the fourth column lists the strongest X-class flare in an AR (GOES class, date and UT time). The 5th column shows the corresponding Flare Index \citep[FI,][]{Abramenko2005}, which was derived by summing the GOES-class of all flares observed in an AR during its passage across the solar disk, $\tau$, and then normalizing the total by $\tau$. Further scaling was applied so that an AR with one C1.0 (X1.0) flare per day has the flare index FI=1.0 (100). The 6th column lists the AR compliance with the empirical laws of the global dynamo for the bipolar ARs. In the headline, the "H" stands for the Hale polarity law, the "J" stands for the Joy's and the "L" for the dominance of the leading spot rule. Below, in lines, the "Y" stands for "Yes" - adhering to the law, and the "N" stands for "No" - violation of the law.  Multipolar ARs are marked with the "M"-symbol and ARs with a strong dominating $\delta$-structure are marked with the symbol "$\delta$".  The magneto-morphological class is shown in the 7th column. And the last column shows the Hale class\footnote{Hale classification data were taken from the following online sources: \url{https://solarmonitor.org} and   \url{http://solarcyclescience.com}.} of an AR determined prior to the flare. As we mentioned above, the majority of ARs (70  out of 79) possessed a $\delta$-structure. 

\begin{figure*}
	\includegraphics[width=\columnwidth]{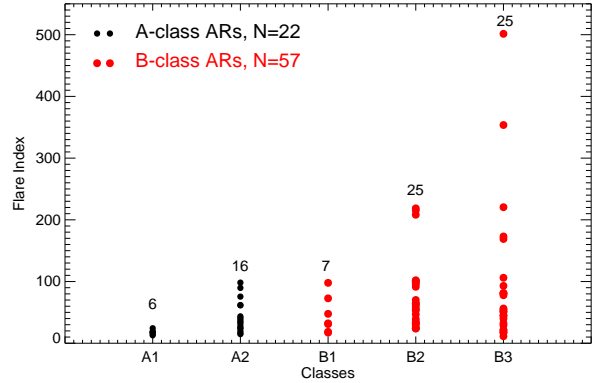}
	\caption{\sf Distribution of 79 ARs with strong flares (>X1.0) over the magneto-morphological classes from A1 to B3. Each AR is marked by a circle (black for regular ARs of A-classes and red for irregular ARs of B-classes). The vertical axis shows the flare index, FI, of each AR. For each class, numbers denote the number of cases. Strongest flares occur (FI>100) only in ARs of classes B2 and B3.  
	}
	\label{fig-6}  
\end{figure*}

The MMC-class distribution of the analyzed ARs is shown in Figure \ref{fig-6}. The majority of the X-class flare ARs (72\%) are of B-class. Also, the AR capability to produce intense flares tends to increase with the MMC-class changing from A1 to B3, i.e., with an enhanced complexity caused by increasing contribution/influence of the turbulent component of the dynamo. (The only deviation is a transition from A2 to B1: the A2-ARs are more numerous than the B1-ARs. Apparently a presence of even small $\delta$-structure is more important for strong flaring than the overall rotation or inclination of the flux tube.) In general, ARs of class A display rather low flare activity, with the most regular ones (A1-class) being the least active and they are small in numbers. ARs with even small $\delta$-structure present (class A2) are more prone to strong flaring, they are more numerous and display higher flare index as compared to the A1-class ARs. Similar dynamics is seen within the B-class ARs; the strongest flares occur in ARs of class B2 and B3, i.e., those ARs that are most affected by the turbulent dynamo.

We would like to emphasize that B-class ARs constitute about 25-30\% of all ARs regardless their flaring activity \citep{Abramenko2018}, whereas their fraction increases up to 72\% when we consider only those ARs with strong flares (>X1.0). This suggests that the occurrence of the most powerful flares is associated with those magnetic configurations for which turbulent dynamo in the convective zone contributed substantially into the generation of their flux.  And the larger the contribution of the turbulent dynamo action, the stronger the flaring potential of the resulting magnetic structure.

\section{Distribution of strong-flaring ARs along a cycle}

\begin{figure*}
	\includegraphics[width=\columnwidth]{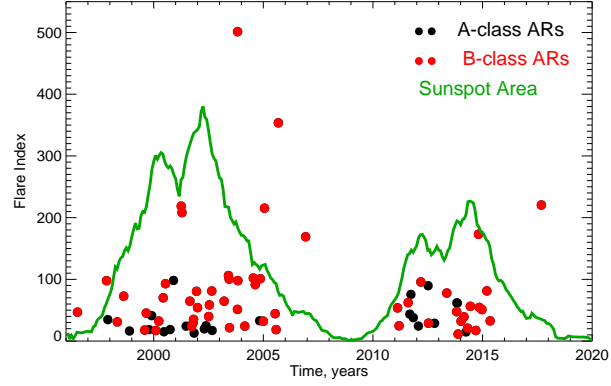}
	\caption{\sf Time distribution of active regions with strong flares (>X1.0) along the two solar cycles. For each AR, the flare index FI is shown along the vertical axis. Regular (irregular) ARs are marked with black (red) circles. The green line shows the total sunspot area smoothed over 13 months; RGO  and USAF/NOAA data available at http://solarcyclescience.com/activeregions.html  were used. Strongest flares occur (FI>100)  only during the second maximum and descending phase of a cycle. 
	  }
	\label{fig-7}  
\end{figure*}

\begin{figure*}
	\includegraphics[width=\columnwidth]{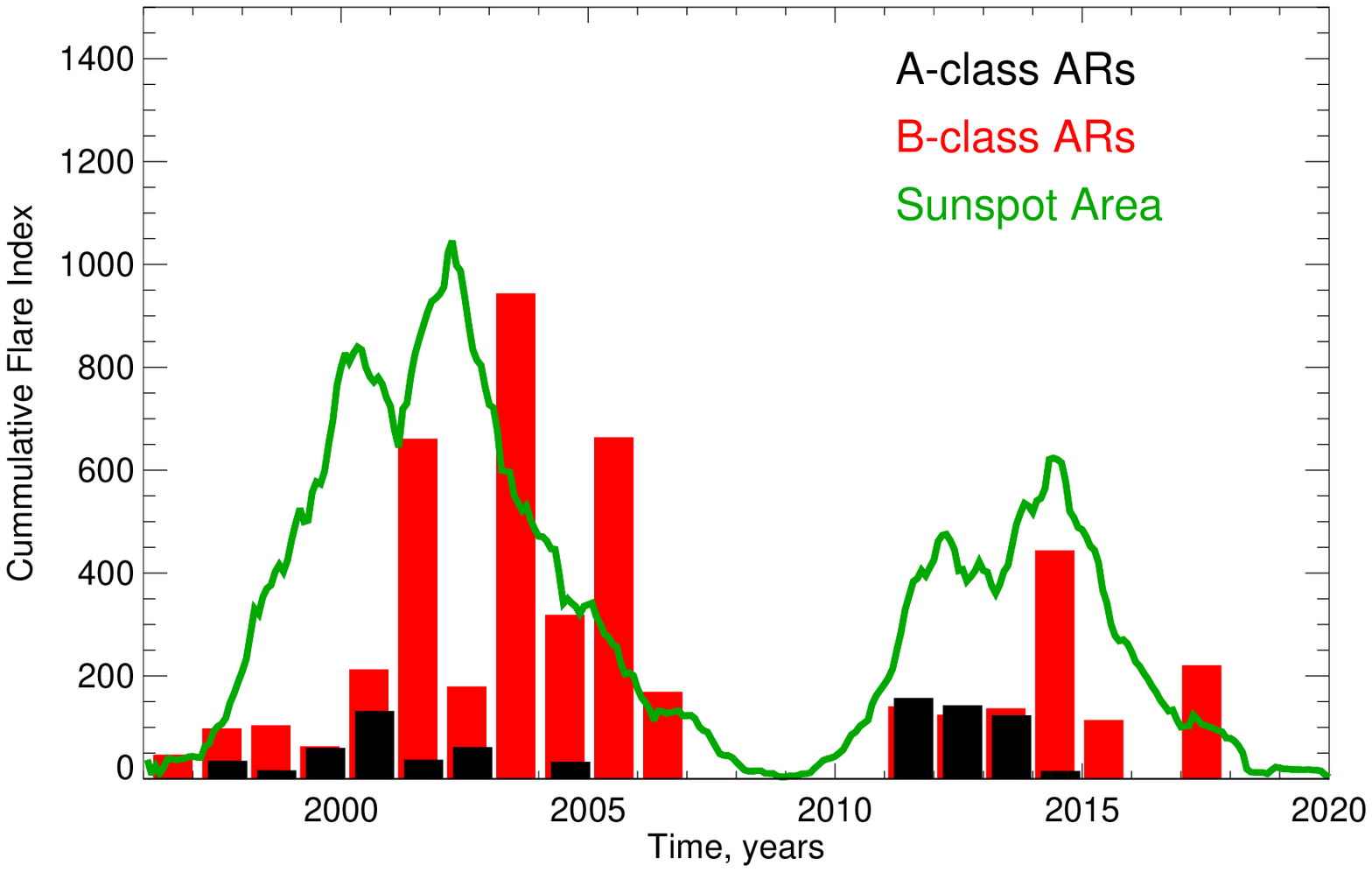}
	\caption{\sf Cumulative over a year flare index of regular (black) and irregular (red) ARs during the two cycles. Regular ARs (a product of the global dynamo) tend to contribute into strong flare production on the rising phase and during the first maximum of the cycle, whereas irregular ARs produce strong flares through the entire cycle and considerably enhance their activity during the second maximum and descending phase. Notations are the same as in Figure \ref{fig-7}. 
	  }
	\label{fig-8}  
\end{figure*}

The cycle dependence in the appearance of flare-productive ARs is shown in Figures \ref{fig-7} and \ref{fig-8} where the AR flare index is plotted  against the AR observation time. We also plot time variations of sunspot area using Royal Greenwich Observatory (RGO) and US Air Force, National Oceanic and Atmospheric Administration (USAF NOAA) sunspot data\footnote{\url{http://solarcyclescience.com/activeregions.html}} smoothed with a 13-month averaging window. Regular A-class ARs are shown with black circles, and the B-class AR are shown with red circles. B-class ARs are more numerous and they appear throughout the entire cycle. The A-class ARs are mostly concentrated at the rising phase and the first maximum of each solar circle. 
There are no A-class ARs with the flare index above $\approx$100 level, so that data for irregular ARs only appear above this level. 

To make this tendency more prominent, we calculated a yearly cumulative flare index as a sum of flare indices of all ARs of a given class (Figure \ref{fig-8}). Flare-productive A-class ARs, which we hypothesize is a product of the global dynamo, tend to appear and contribute into the flaring at the rising phase and during the first maximum of a cycle, when the global toroidal field is expected to be strong. However, during the second maximum and the declining phase of a solar cycle, when the toroidal field weakens \citep{Charbonneau2020}, the irregular ARs become the main producers of powerful episodes of solar activity. 

\section{Concluding remarks}  

We introduced a magneto-morphological classification (MMC) of ARs in order to better describe possible contribution of turbulent dynamo into the formation of ARs. Comparing the MMC class of active region and their flare productivity over solar cycles 23 and 24 allowed us to conclude the following. 

\noindent 1) Out of all ARs that produced X-class flares, 72\% are the ARs-``violators'', i.e., non-compliant with (at least one of) the empirical laws of the global dynamo (Hale polarity law, Joy's law, and the leading spot prevalence rule), while these ARs constitute only 25-30\% of all observed 1494 ARs \citep{Abramenko2018}. 
Thus,  the strongest fluctuations in the time domain (flares) are 
statistically related to the strongest distortions
 in the space domain, which is one of the key properties of a non-linear dynamical dissipative system. The inference is in a favour of a viewpoint that the solar dynamo is one of such systems.   

\noindent 2) The time distribution of flaring ARs over a solar cycle indicates that the regular A-class ARs contribute to solar activity mostly during the rising phase of a cycle and its first maximum, whereas irregular B-class ARs are more distributed in the decline phase and they are dominating source of solar flares during the second maximum and the declining phase. The rising phase of the dynamo wave is when the toroidal component of the magnetic field is strongest so that we observe both regular and irregular flaring ARs. As the dynamo wave proceeds, the toroidal component weakens and the turbulent component of the dynamo becomes more pronounced thus notably distorting emerging toroidal flux tubes and leading to appearance of strong fluctuations in the spatial domain (i.e., irregular ARs).  

Our analysis showed that the majority of ARs that produced X-class flares do not follow the laws of the global mean-field dynamo and possess an irregular magnetic structure. Therefore, large temporal and spatial fluctuations in the solar dynamo are not rare and indicate on the existence of the turbulent component of the dynamo on scales of ARs. Thus the gap between the large-scale dynamo that generates the global poloidal and toroidal fields and the small-scale turbulent dynamo that is responsible for quiet sun magnetic fields may be filled by the turbulent dynamo acting on mid-scales throughout the convective zone. Then one might expect a continuous spectrum of turbulent magnetic fields and energy on a large range of spatial scales. The presumed continuous spectrum is a natural property of a turbulent medium \citep{Monin1975}. Note that existence of a continuous temporal spectrum of solar activity was recently demonstrated by \citet{Frick2020} based on time variations of the total sunspot area. The continuous spectrum implies that magnetic energy is generated not only on the largest scales, but also on a wide range of intermediate scales of the turbulent intermittent convective zone. The entire process works as a whole with a continuous energy exchange between the scales. The concept was suggested in early theoretical studies by \citet{Kazantsev1968} and \citet{Zeldovich1987}. More recently the existence of the turbulent component of dynamo follows from theoretical considerations and numerical simulations in both time variations \citep[][]{Sokoloff2010,Olemskoy2013,Passos2014,Karak2017,Schussler2018} 
and spatial properties  \citep[][]{Nelson2013} of solar activity. However, observational evidence of the continuous spectrum, especially in the spatial domain, is not strong so far mainly because the dynamo action is hidden below the photosphere. Publications in this field are rather scanty \citep{Sokoloff2015}.

In conclusion, it is worth to mention that the explored here four essential properties which characterise the regular ARs (bipolarity and three empirical laws)  do not cover the entire list of such properties. For example, the hemispheric sign preference rule of helicity complies for the majority of ARs \citep[][and references herein]{Pevtsov2014}. This parameter undoubtedly deserves an extended analysis in future. Note that \citet{LaBonte2007} investigates ARs of the 23rd solar cycle and found that for X-flaring ARs, the hemispheric sign preference rule tends to be obscured due to intrinsic helicity injection of opposite sign.  A recent study by \citet{Park2021} of the hemispheric sign preference rule for ARs of the 24th solar cycle demonstrated that in heliographic areas where the ARs with strong flares occurred, the degree of compliance  of this rule is lowered. They argued that below the photosphere there should be localized volumes of enhanced turbulence, where vigorous turbulent plasma motions affect the shape and future flare capability of some flux tubes while they are rising to the surface. This inference is in agreement with the suggested above concept on the role of the turbulent component of dynamo in the AR formation and flaring.


\begin{table*}
	\caption{ARs with X-class flares in the 23rd cycle}
	\label{tab2}
\begin{tabular}{ l c c l r l l l } 
	\hline
	NOAA  & Start & End & Max. flare (UT) & FI & H J L$^a$ & MMC class  & Hale class	\\
	\hline
	07978   &   1996.07.07  &  1996.07.13   &   X2.6 1996.07.09 (09:01)    &    46.85   &    M$^{\star}$  &   B2  &   $\beta \gamma \delta$	\\
	08100	&	1997.10.27	&	2997.11.09	&	X9.4 1997.11.06 (11:49)	   &	97.85	&	YYN	&	B1	&	$\beta \gamma \delta$	\\
	08113	&	1997.11.26	&	1997.12.09	&	X2.6 1997.11.27 (12:59)    &	36.00	&	YYY	&	A2	&	$\beta \gamma \delta$	\\
	08210	&	1998.04.25	&	1998.05.08	&	X1.1 1998.05.02 (13:31)	   &	31.11	&	NYN	&	B1	&	$\beta \gamma \delta$	\\
	08307	&	1998.08.19	&	1998.09.02	&	X4.9 1998.08.18 (22:10)	   &	72.74	&	YYN	&	B1	&	$\beta \delta$	\\
	08384	&	1998.11.09	&	1998.11.23	&	X2.2 1998.11.23 (06:28)    &	16.29	&	YYY	&	A2	&	$\beta \delta$	\\
	08647*	&	1999.07.25	&	1999.08.06	&	X1.4 1999.08.02 (21:18)    &	17.85	&	M	&	B3 	&	$\beta \gamma$	\\
	08674   &   1999.08.20  &  1999.09.02   &   X1.1 1999.08.28 (17:52)    &    45.33   &   M   &   B3  &   $\beta \gamma \delta$	\\
	08731   &   1999.10.10  &  1999.10.23   &   X1.8 1999.10.14 (08:54)    &    18.52   &   YYY &   A1  &   $\beta \gamma $		 \\
	08771	&	1999.11.18	&	1999.11.29	&	X1.4 1999.11.27 (12:05) 	&	36.81	&	YYY	&	A2	&	$\beta \gamma \delta$	\\
	08858	&	2000.02.03	&	2000.02.16	&	X1.2 2000.02.05 (19:17) 	&	17.18	&	M	&	B3  &	$\beta $	\\
	08910	&	2000.03.12	&	2000.03.25	&	X1.8 2000.03.24 (07:41) 	&	32.37	&	M	&	B3  &	$\beta \gamma \delta$	\\
	09026	&	2000.06.01	&	2000.06.14	&	X2.3 2000.06.06 (14:58) 	&	70.07	&$\delta$$^{\star}$&   B2  &   $\beta \gamma \delta$	\\
	09033	&	2000.06.05	&	2000.06.19	&	X1.0 2000.06.18 (01:52)     &	15.04	&	YYY	&	A1  &  $\beta \gamma $		 \\
	09077	&	2000.07.07	&	2000.07.21	&	X5.7 2000.07.14 (10:03)	    &	92.96	&	M	&	B3  &  $\beta \gamma \delta$ \\
	09169	&	2000.09.18	&	2000.10.01	&	X1.2 2000.09.30 (23:13) 	&	18.67	&	YYY	&	A1	 &	$\beta \gamma \delta$	\\
	09236	&	2000.11.18	&	2000.12.01	&	X4.0 2000.11.26 (16:34)     &	98.22	&	YYY	&	A2    &	$\beta \gamma\delta $   \\
	09393   &  2001.03.23   &  2001.04.04   &   X20. 2001.04.02 (21:32)     &  218.59   &   M   &   B2    & $\beta \gamma\delta $   \\
	09415   &  2001.04.03   &  2001.04.15   &   X14. 2001.04.15 (13:19)     &  208.20   &$\delta$&  B2    & $\beta \gamma\delta $   \\
	09511   &  2001.06.20   &  2001.06.30   &   X1.2 2001.06.23 (04:02)     &   24.17   &   YYY &   A2    & $\beta \gamma\delta $  \\
	09591   &  2001.08.22   &  2001.09.03   &   X5.6 2001.08.25 (16:23)     &   64.59   &$\delta$&  B2    & $\beta \gamma\delta $  \\
	09632   &  2001.09.20   &  2001.10.02   &   X2.6 2001.09.24 (09:32)     &   23.78   &$\delta$ &   B2    & $\beta \gamma\delta $ \\  
	09661   &  2001.10.11   &  2001.10.23   &   X1.6 2001.10.19 (00:47)     &   30.07   &$\delta$ &   B2    & $\beta \gamma\delta $  \\
	09672   &  2001.10.18   &  2001.10.30   &   X1.3 2001.10.25 (14:42)     &   35.11   &$\delta$ &   B2    & $\beta \gamma\delta $ \\
	09684   &  2001.10.28   &  2001.11.09   &   X1.0 2001.11.04 (16:03)     &   12.89   &   YYY &   A1    & $\beta \gamma $		 \\
	09733   &  2001.12.08   &  2001.12.20   &   X6.2 2001.12.13 (14:20)     &   80.52   &   M &   B3    & $\beta \gamma\delta $	\\
	09906   &   2002.04.11  &  2002.04.21   &   X1.5 2002.04.21 (00:43)     &   19.72   &   YYY &   A2   & $\beta \gamma\delta $   \\
	09961   &   2002.05.19  &  2002.06.01   &   X2.1 2002.05.20 (15:21)     &   24.74   &   YYY &   A2   & $\beta \gamma\delta $   \\ 
	10017   &   2002.06.28  &  2002.07.05   &   X1.5 2002.07.03 (02:13)     &   39.75   &   M   &   B3   & $\beta \gamma\delta $   \\
	10030   &   2002.07.09  &  2002.07.22   &   X3.0 2002.07.15 (20:08)     &   58.59   &   M   &   B2   & $\beta \gamma\delta $   \\
	10039   &   2002.07.22  &  2002.08.04   &   X4.8 2002.07.23 (00:18)     &   54.15   &   M   &   B3   & $\beta \gamma\delta $   \\
	10069   &   2002.08.11  &  2002.08.24   &   X3.1 2002.08.24 (00:49)     &   81.11   &   M   &   B3   & $\beta \gamma\delta $   \\
	10095   &   2002.08.29  &  2002.09.10   &   X1.5 2002.08.30 (12:47)     &   17.11   &   YYY &   A1   & $\beta \gamma $		 \\
	10314	&	2003.03.14	&	2003.03.21	&	X1.5 2003.03.17 (18:50) 	&	64.62	&	M  	&	B2 	&$\beta \gamma \delta$		\\
	10365	&	2003.05.20	&	2003.06.03	&	X3.6 2003.05.28 (00:17)	    &  106.11	&	M	&	B3 	&$\beta \gamma \delta$		\\
	10375	&	2003.06.01	&	2003.06.14	&	X1.7 2003.06.09 (21:31)	    &  100.44	&	M	&	B2  &$\beta \gamma \delta$		\\
	10386   &   2003.06.15  &   2003.06 25  &   X1.3 2003.06.15 (23:25)     &   21.64   &   M   &   B3  & $\beta \gamma \delta$		\\
	10484	&	2003.10.18	&	2003.10.31	&	X1.2 2003.10.26 (17:21)	    &	51.33	&	M	&	B3 	&$\beta \gamma \delta$		\\
	10486	&	2003.10.22	&	2003.11.05	&	X17  2003.10.28 (09:51)	    &  501.41	&	M	&	B3 	&$\beta \gamma \delta$		\\
	10488	&	2003.10.27	&	2003.11.04	&	X3.9 2003.11.03 (09:43)	    &	98.00	&	M	&	B2	&$\beta \gamma \delta$		\\
	10564   &   2004.02.21  &   2004.03.02  &   X1.1 2004.02.26 (02:03)     &   24.22   &   M   &   B2  &$\beta \gamma \delta$		\\
	10649	&	2004.07.12	&	2004.07.25	&	X3.6 2004.07.16 (07:51)	    &  102.07	&$\delta$&	B2	&$\beta \gamma \delta$		\\
	10656	&	2004.08.06	&	2004.08.19	&	X1.8 2004.08.18 (17:29)  	&	91.48	&	M 	&	B2  &$\beta \gamma \delta$		\\
	10691	&	2004.10.24	&	2004.11.05	&	X1.2 2004.10.30 (11:38)	    &	32.89	&	YYY	&	A2	&$\beta \gamma \delta$		\\
	10696	&	2004.11.02	&	2004.11.12	&	X2.5 2004.11.10 (01:59)	    &	101.00	&	M	&	B2	&$\beta \gamma \delta$		\\
	10715   &   2004.12.28  &  2005.01.09   &   X1.7 2005.01.01 (00:01)     &    32.20  &   NYY &   B1  &$\beta \gamma \delta$		\\
	10720   &   2005.01.11  &  2005.01.21   &   X7.1 2005.01.20 (06:36)     &   215.27  &$\delta$&   B2  &$\beta \delta$		\\
	10786   &   2005.07.02  &  2005.07.14   &   X1.2 2005.07.14 (10:16)     &    44.22  &   M   &   B3  & $\beta \gamma \delta$		\\
	10792   &   2005.07.29  &  2005.08.09   &   X1.3 2005.07.30 (06:17)     &    18.52  &   NNN &   B1  & $\beta \gamma \delta$		\\
	10808   &   2005.09.07  &  2005.09.19   &   X17  2005.09.07 (17:17)     &   353.63  &   M   &   B3  & $\beta \gamma \delta$		\\ 
	10930	&	2006.12.06	&	2006.12.18	&	X9.0 2006.12.05 (10:18)   	&	168.96	&   M   &	B3	&$\beta \gamma \delta$		\\
	
	\hline
	\multicolumn{8}{l}{\footnotesize{$^a$ - for bipolar ARs: "H" stands for the Hale polarity law, the "J" for the Joy's law, the "L" for the leader prevalence rule.}}\\
	\multicolumn{8}{l}{\footnotesize{Below, in lines, for bipolar ARs: the "Y" stands for "Yes" - adhering to the law, and the "N" stands for "No" - violation of the law. }}\\
	\multicolumn{8}{l}{\footnotesize{$^{\star}$ - "M" marks multipolar ARs, "$\delta$" marks ARs with tight strong dominating $\delta$-structure. 		 }}\\
	\multicolumn{8}{l}{\footnotesize{$^*$ - the strongest flare occurred between two ARs. }}\\

    \end{tabular}	
\end{table*}

\begin{table*}
	\caption{ARs with X-class flares in the 24th cycle}
	\label{tab3}
	\begin{tabular}{ l c c l r l l l } 
		\hline
		NOAA  & Start & End & Max. flare (UT) & FI & H J L & MMC class  & Hale class	\\
		\hline
			11158	&	2011.02.11	&	2011.02.21	&	X2.2 2011.02.15 (01:44) 	&	53.72	&	M	&	B2  &$\beta \gamma \delta$		\\
			11166	&	2011.03.03	&	2011.03.16	&	X1.5 2011.03.09 (23:13) 	&	24.74	&	M	&	B2  &$\beta \gamma \delta$		\\
			11263	&	2011.07.28	&	2011.08.11	&	X6.9 2011.08.09 (07:48) 	&	62.30	&	M	&	B2  &$\beta \gamma \delta$		\\
			11283	&	2011.08.30	&	2011.09.12	&	X2.1 2011.09.06 (22:12)	    &	43.55	&	YYY	&	A2	&$\beta \gamma \delta$		\\
			11302	&	2011.09.22	&	2011.10.05	&	X1.9 2011.09.24 (09:21)	    &	75.55	&	YYY	&	A2	&$\beta \gamma \delta$		\\
			11339	&	2011.11.01	&	2011.11.15	&	X1.9 2011.11.03 (20:16)	    &	37.92	&	YYY	&	A2	&$\beta \gamma \delta$		\\
			11402	&	2012.01.14  &	2012.01.28	&	X1.7 2012.01.27 (17:37)	    &	24.29	&	YYY	&	A1	&$\beta \gamma $		\\
			11429	&	2012.03.03	&	2012.03.16	&	X5.4 2012.03.07 (00:02)	    &	95.78	&	M	&	B2  &$\beta \gamma \delta$		\\
			11515   &  2012.06.27   &  2012.07.09   &   X1.1 2012.07.06 (23:01)     &    89.70  &   YYY &   A2  & $\beta \gamma \delta$		\\
			11520	&  2012.07.07	&  2012.07.19	&	X1.4 2012.07.12 (15:37)	    &	 28.96	&	M	&	B3  &$\beta \gamma \delta$		\\
			11598	&	2012.10.21	&  2012.11.02	&	X1.8 2012.10.23 (03:13)	    &	28.74	&	YYY	&	A2	&$\beta \delta$		\\
			11748	&	2013.05.13	&  2013.05.26	&	X3.2 2013.05.13 (23:59)	    &	77.78	&	M	&	B3  &$\beta \gamma \delta$		\\
			11875	&	2013.10.17	&  2013.10.30	&	X2.3 2013.10.29 (21:42)     &	61.85	&	YYY	&	A2	&$\beta \gamma \delta$		\\
			11882	&	2013.10.25	&  2013.11.06	&	X2.1 2013.10.25 (14:51)	    &	47.63	&	YNN	&	B1 &$\beta \gamma \delta$		\\
			11890	&	2013.11.03	&  2013.11.16	&	X3.3 2013.11.05 (22:07)	    &	61.41	&	YYY	&	A2	&$\beta \gamma \delta$		\\
			11893	&	2013.11.09  &  2013.11.21	&	X1.0 2013.11.19 (10:14)     &	11.33	&	M	&	B3  &$\beta \delta$		\\
			11944   &   2014.01.01  &  2014.01.14   &   X1.2 2014.01.07 (18:04)     &   32.15   &   M &   B2  & $\beta \gamma $		\\
			11990	&	2014.02.25	&  2014.03.10	&	X4.9 2014.02.25 (00:39)	    &	39.26	&$\delta$&	B2	&$\beta \delta$		\\
			12017	&	2014.03.22	&  2014.04.03	&	X1.0 2014.03.29 (17:35)     &	14.81	&	YYY	&	A2	&$\beta \delta$		\\
			12035   &   2014.04.11  &  2014.04.24   &   X1.3 2014.04.25 (00:27)     &   20.84   &   M   &   B3  & $\beta \gamma $		\\
			12087	&	2014.06.10	&   2014.06.23	&	X2.2 2014.06.10 (11:36)     &	56.30	&	M	&	B3  &$\beta \delta$		\\
			12158	&   2014.09.05	&	2014.09.18	&	X1.6 2014.09.10 (17:21) 	&	17.18	&	NNY	&	B1	&$\beta \gamma \delta$		\\
			12192	&	2014.10.18	&   2014.10.31	&	X3.1 2014.10.24 (21:07)     &	173.04	&	M	&	B3  &$\beta \gamma \delta$		\\
			12205	&	2014.11.04	&	2014.11.17	&	X1.6 2014.11.07 (16:53)	    &	54.52	&	M	&	B2	&$\beta \gamma \delta$		\\
			12242	&	2014.12.14	&   2014.12.24	&	X1.8 2014.12.20 (00:11)	    &	51.00	&	M	&	B3  &$\beta \gamma \delta$		\\
			12297	&	2015.03.07	&   2015.03.20	&	X2.1 2015.03.11 (16:11)	    &	81.26	&	M	&	B3 &$\beta \gamma \delta$		\\
			12339   &  2015.05.05   &  2015.05.17   &   X2.7 2015.05.05 (22:11)     &   32.74   &   M   &   B2 & $\beta \gamma$		\\
			12673	&	2017.08.30	&  2017.09.10	&	X9.3 2017-09-06 (11:53)	    &	220.44	&	M	&	B3 &$\beta \gamma \delta$		\\
		\hline
		
		\multicolumn{8}{l}{\footnotesize{Notations are the same as for Table \ref{tab2}.}}\\
		
	\end{tabular}	
\end{table*}

\section*{Acknowledgements}

 I am thankful to anonymous referee whose comments helped much to improve the paper. SDO is a mission for NASA Living With a Star (LWS) program. The SDO/HMI data were provided by the Joint Science Operation Center (JSOC). The study was supported by Russian Science Foundation grant 18-12-00131.

\section{Data availability}

The MDI and HMI data that support the findings of this study are available
in the JSOC (http://jsoc.stanford.edu/) and can be accessed under
open for all data policy.

Derived data products supporting the findings of this study
are available from the corresponding author VA on request.







\bsp	
\label{lastpage}
\end{document}